\documentclass[amsmath,amssymb,superscriptaddress,floatfix,reprint,prl]{revtex4-1}
\pdfoutput=1
\usepackage{graphicx}
\usepackage{hyperref}
\usepackage{dcolumn}
\usepackage{gensymb}
\usepackage[caption=false]{subfig}

\begin{document}

\title{Evidence for Astrophysical Muon Neutrinos from the Northern Sky with IceCube}
% Author list as of 20150508

\affiliation{III. Physikalisches Institut, RWTH Aachen University, D-52056 Aachen, Germany}
\affiliation{School of Chemistry \& Physics, University of Adelaide, Adelaide SA, 5005 Australia}
\affiliation{Dept.~of Physics and Astronomy, University of Alaska Anchorage, 3211 Providence Dr., Anchorage, AK 99508, USA}
\affiliation{CTSPS, Clark-Atlanta University, Atlanta, GA 30314, USA}
\affiliation{School of Physics and Center for Relativistic Astrophysics, Georgia Institute of Technology, Atlanta, GA 30332, USA}
\affiliation{Dept.~of Physics, Southern University, Baton Rouge, LA 70813, USA}
\affiliation{Dept.~of Physics, University of California, Berkeley, CA 94720, USA}
\affiliation{Lawrence Berkeley National Laboratory, Berkeley, CA 94720, USA}
\affiliation{Institut f\"ur Physik, Humboldt-Universit\"at zu Berlin, D-12489 Berlin, Germany}
\affiliation{Fakult\"at f\"ur Physik \& Astronomie, Ruhr-Universit\"at Bochum, D-44780 Bochum, Germany}
\affiliation{Physikalisches Institut, Universit\"at Bonn, Nussallee 12, D-53115 Bonn, Germany}
\affiliation{Universit\'e Libre de Bruxelles, Science Faculty CP230, B-1050 Brussels, Belgium}
\affiliation{Vrije Universiteit Brussel, Dienst ELEM, B-1050 Brussels, Belgium}
\affiliation{Dept.~of Physics, Chiba University, Chiba 263-8522, Japan}
\affiliation{Dept.~of Physics and Astronomy, University of Canterbury, Private Bag 4800, Christchurch, New Zealand}
\affiliation{Dept.~of Physics, University of Maryland, College Park, MD 20742, USA}
\affiliation{Dept.~of Physics and Center for Cosmology and Astro-Particle Physics, Ohio State University, Columbus, OH 43210, USA}
\affiliation{Dept.~of Astronomy, Ohio State University, Columbus, OH 43210, USA}
\affiliation{Niels Bohr Institute, University of Copenhagen, DK-2100 Copenhagen, Denmark}
\affiliation{Dept.~of Physics, TU Dortmund University, D-44221 Dortmund, Germany}
\affiliation{Dept.~of Physics and Astronomy, Michigan State University, East Lansing, MI 48824, USA}
\affiliation{Dept.~of Physics, University of Alberta, Edmonton, Alberta, Canada T6G 2E1}
\affiliation{Erlangen Centre for Astroparticle Physics, Friedrich-Alexander-Universit\"at Erlangen-N\"urnberg, D-91058 Erlangen, Germany}
\affiliation{D\'epartement de physique nucl\'eaire et corpusculaire, Universit\'e de Gen\`eve, CH-1211 Gen\`eve, Switzerland}
\affiliation{Dept.~of Physics and Astronomy, University of Gent, B-9000 Gent, Belgium}
\affiliation{Dept.~of Physics and Astronomy, University of California, Irvine, CA 92697, USA}
\affiliation{Dept.~of Physics and Astronomy, University of Kansas, Lawrence, KS 66045, USA}
\affiliation{Dept.~of Astronomy, University of Wisconsin, Madison, WI 53706, USA}
\affiliation{Dept.~of Physics and Wisconsin IceCube Particle Astrophysics Center, University of Wisconsin, Madison, WI 53706, USA}
\affiliation{Institute of Physics, University of Mainz, Staudinger Weg 7, D-55099 Mainz, Germany}
\affiliation{Universit\'e de Mons, 7000 Mons, Belgium}
\affiliation{Technische Universit\"at M\"unchen, D-85748 Garching, Germany}
\affiliation{Bartol Research Institute and Dept.~of Physics and Astronomy, University of Delaware, Newark, DE 19716, USA}
\affiliation{Department of Physics, Yale University, New Haven, CT 06520, USA}
\affiliation{Dept.~of Physics, University of Oxford, 1 Keble Road, Oxford OX1 3NP, UK}
\affiliation{Dept.~of Physics, Drexel University, 3141 Chestnut Street, Philadelphia, PA 19104, USA}
\affiliation{Physics Department, South Dakota School of Mines and Technology, Rapid City, SD 57701, USA}
\affiliation{Dept.~of Physics, University of Wisconsin, River Falls, WI 54022, USA}
\affiliation{Oskar Klein Centre and Dept.~of Physics, Stockholm University, SE-10691 Stockholm, Sweden}
\affiliation{Dept.~of Physics and Astronomy, Stony Brook University, Stony Brook, NY 11794-3800, USA}
\affiliation{Dept.~of Physics, Sungkyunkwan University, Suwon 440-746, Korea}
\affiliation{Dept.~of Physics, University of Toronto, Toronto, Ontario, Canada, M5S 1A7}
\affiliation{Dept.~of Physics and Astronomy, University of Alabama, Tuscaloosa, AL 35487, USA}
\affiliation{Dept.~of Astronomy and Astrophysics, Pennsylvania State University, University Park, PA 16802, USA}
\affiliation{Dept.~of Physics, Pennsylvania State University, University Park, PA 16802, USA}
\affiliation{Dept.~of Physics and Astronomy, Uppsala University, Box 516, S-75120 Uppsala, Sweden}
\affiliation{Dept.~of Physics, University of Wuppertal, D-42119 Wuppertal, Germany}
\affiliation{DESY, D-15735 Zeuthen, Germany}

\author{M.~G.~Aartsen}
\affiliation{School of Chemistry \& Physics, University of Adelaide, Adelaide SA, 5005 Australia}
\author{K.~Abraham}
\affiliation{Technische Universit\"at M\"unchen, D-85748 Garching, Germany}
\author{M.~Ackermann}
\affiliation{DESY, D-15735 Zeuthen, Germany}
\author{J.~Adams}
\affiliation{Dept.~of Physics and Astronomy, University of Canterbury, Private Bag 4800, Christchurch, New Zealand}
\author{J.~A.~Aguilar}
\affiliation{Universit\'e Libre de Bruxelles, Science Faculty CP230, B-1050 Brussels, Belgium}
\author{M.~Ahlers}
\affiliation{Dept.~of Physics and Wisconsin IceCube Particle Astrophysics Center, University of Wisconsin, Madison, WI 53706, USA}
\author{M.~Ahrens}
\affiliation{Oskar Klein Centre and Dept.~of Physics, Stockholm University, SE-10691 Stockholm, Sweden}
\author{D.~Altmann}
\affiliation{Erlangen Centre for Astroparticle Physics, Friedrich-Alexander-Universit\"at Erlangen-N\"urnberg, D-91058 Erlangen, Germany}
\author{T.~Anderson}
\affiliation{Dept.~of Physics, Pennsylvania State University, University Park, PA 16802, USA}
\author{M.~Archinger}
\affiliation{Institute of Physics, University of Mainz, Staudinger Weg 7, D-55099 Mainz, Germany}
% ...

\date{\today}

\collaboration{IceCube Collaboration}
\noaffiliation

\begin{abstract}
Results from the IceCube Neutrino Observatory have recently provided compelling evidence for the existence of a high energy astrophysical neutrino flux utilizing a dominantly Southern Hemisphere dataset consisting primarily of $\nu_e$ and $\nu_\tau$ charged current and neutral current (cascade) neutrino interactions. 
In the analysis presented here, a data sample of approximately 35,000 muon neutrinos from the Northern sky was extracted from data taken during 659.5 days of livetime recorded between May 2010 and May 2012. 
While this sample is composed primarily of neutrinos produced by cosmic ray interactions in the Earth's atmosphere, the highest energy events are inconsistent with a hypothesis of solely terrestrial origin at $3.7 \sigma$ significance. 
These neutrinos can, however, be explained by an astrophysical flux per neutrino flavor at a level of $\Phi(E_{\nu}) = 9.9^{+3.9}_{-3.4} \times 10^{-19}\, \mathrm{GeV}^{-1}\, \mathrm{cm}^{-2}\, \mathrm{sr}^{-1}\, \mathrm{s}^{-1} \left({E_{\nu} \over 100 \mathrm{TeV}}\right)^{-2}$, consistent with IceCube's Southern Hemisphere dominated result. 
Additionally, a fit for an astrophysical flux with an arbitrary spectral index was performed.
We find a spectral index of $2.2^{+0.2}_{-0.2}$, which is also in good agreement with the Southern Hemisphere result. 
\end{abstract}

\maketitle

%\textit{Introduction} ---
%Connection between neutrinos and cosmic rays
The nature of the objects and the mechanisms which accelerate cosmic rays pose major open questions in current astrophysics, which may, in part, be answered by observations of high energy neutrinos. 
At high energies, the majority of cosmic rays are protons or atomic nuclei, and their interaction with other matter or radiation is known to produce neutrinos \cite{ref:stecker1979}. 
If this happens near the source of the cosmic rays, the neutrinos, which---unlike the charged cosmic rays---can travel undeflected through the magnetic fields of deep space, can point back to these sources. 

IceCube is a detector constructed at depths between 1.5 km and 2.5 km in glacial ice at the South Pole, instrumenting about a cubic kilometer of volume with optical sensors \cite{ref:DAQ_paper}. 
This forms a Cherenkov detector for the light produced when neutrinos interact and generate secondary charged particles. 
These interactions give rise to two characteristic event topologies: linear `tracks,' produced by long-range muons emitting light as they travel, and near-spherical `cascades,' from the more point-like light emission of electromagnetic and hadronic particle showers which terminate in ice after small distances compared to the instrumentation density of the detector~\cite{ref:IC_perf}. 

%The HESE result
One effective method for identifying neutrino interactions is to look for events which show no sign of light emission when entering the detector boundary. 
These are referred to as `starting' events. 
A recent IceCube study using this technique \cite{ref:HESE2} has determined that astrophysical neutrinos at high energies do exist, and that their flux is broadly compatible with existing models \cite{ref:LoebWaxman,ref:Stecker2005,ref:WB_GRB}. 
%Contrast between this analysis and HESE
While such starting events provide good evidence for an astrophysical neutrino flux, they do not sample all components of the expected flux equally well. 
Due to absorption in the Earth, few neutrinos are observed from the Northern sky, and few of the observed events are identifiably $\nu_\mu$. 
This analysis seeks to observe more of these particular types of events by relaxing the requirement that events begin inside the detector to permit the use of the long muon range to achieve a larger effective volume. 
Events are then selected based on the event topology of muons produced from $\nu_\mu$ interactions to reduce background contamination. 
In this analysis, as in other IceCube analyses, it is not possible to distinguish neutrinos from antineutrinos, so only the combined flux can be measured. 

%\textit{Analysis} ---
To identify astrophysical muon neutrinos, we must distinguish them both from other types of events in the detector and from other sources of neutrinos. 
The majority of the data recorded by IceCube are produced by muons originating in cosmic ray air showers that penetrate the ice and reach the detector. 
Since this analysis seeks to take advantage of the long muon tracks and cannot depend on observing the neutrino interaction vertex inside the detector, only muons with directions that imply they passed through more material than the maximal expected muon range are selected. 
In this case, part of the distance must have been traversed by a neutrino, which is less prone to interaction.
This analysis accepts therefore only events whose reconstructed zenith angles are greater than 85$\degree$, corresponding to an overburden equivalent to at least 12 km of water. 
The directions of muon events are reconstructed by fitting the hypothesis of a particle moving at the speed of light and emitting Cherenkov radiation to the timing of the observed photons. The fit accounts for the expected delay of the first photon to reach each detector module due to scattering~\cite{ref:reco}.
Rejecting poorly fit events removes both low energy atmospheric muons with poor direction resolution and the less numerous cascade-like events produced by neutrino interactions other than charged-current $\nu_\mu$. 
In addition to the direction of the muon, the other observable of interest is muon energy.
A proxy for the energy is computed by fitting the amount of light expected to be emitted by a template muon to the number of observed photons in each event \cite{ref:energy}
\footnote{
	The performance of the energy reconstruction used in this analysis is shown in Fig. 22 of~\cite{ref:energy} and is discussed in~\cite{ref:supplement}. 
}.
The precision of the energy proxy is limited by the relatively short section of the muon's total track which is observed, and is only loosely connected to the energy of the interacting neutrino since an unknown amount of energy is generally lost before the muon reaches the detector. 
After applying event-quality criteria (which are qualitatively equivalent to those used in earlier studies \cite{ref:IC40,ref:IC59}, with details being given in the online supplement \cite{ref:supplement} and in \cite{ref:thesis}) this yields a highly pure ($99.9\%$) sample of neutrino-induced muon events, with an efficiency of about 24\% for neutrino-induced events from an $E^{-2}$ spectrum. 
This selection still suffers from neutrino absorption in the Earth, resulting in a loss of events at the highest zenith angles and energies. 
This analysis was performed with a blindness criterion such that only 10\% of the experimental data were used in its development, in conjunction with simulated data, to determine the data selection
The full data are used only after the analysis technique had been fixed. 

%Expected fluxes (atmospheric, astrophysical)
Since the astrophysical neutrinos we seek to observe in this study are expected to be produced in conjunction with the cosmic rays~\cite{ref:GW_CR,ref:EngelMann_CR}, they should have a related power-law spectrum of the form $\Phi \propto E^{-\gamma}$, where $\gamma$ should be $\sim 2$. 
For this analysis we take $\gamma=2$ as a benchmark model \cite{ref:WB2013}. 
We also make the further simplifying assumption that the astrophysical flux is isotropic, as would be the case for a signal originating from many distant, individually weak sources. 

Although astrophysical neutrinos are the target of the analysis, the numerous neutrinos produced by cosmic ray air showers must be accounted for. 
Atmospheric neutrinos are usually separated into two groups: those produced by the decays of pions and kaons, referred to as `conventional,' 
and those produced by the decays of heavier mesons, particularly those containing charm quarks, referred to as `prompt'.
Since the conventional atmospheric neutrinos arise from relatively well-understood particle physics and have been measured by a variety of experiments \cite{ref:FrŽjus,ref:SuperK}, there exist several models for this flux~\cite{ref:Honda,ref:Bartol,ref:Fedynitch_atmos}
Here we use the HKKMS07 calculation \cite{ref:Honda}, where the uncertainty of this calculation is estimated by its authors to be less than 10\% at few GeV energies, which is consistent with measurements \cite{ref:SuperK_atm_w_proton}, and is expected to increase with energy to around 25\% at 1 TeV. 
Since this model was designed for relatively low energies (100 MeV-10 TeV) compared to those considered in this analysis ($\sim$100 GeV-100 TeV), it is extended and modified according to the procedure in \cite{ref:IC59} to take into account the input cosmic ray spectrum \cite{ref:H3a} at high energies. 
An important feature of the conventional atmospheric neutrino flux is that the parent mesons may be destroyed by interactions with the medium before decaying and producing neutrinos. 
The energy spectrum is therefore steeper ($\propto E^{-3.7}$) than that of the cosmic rays from which it is produced ($\propto E^{-2.7}$) \cite{ref:GaisserBook}. 
This is then markedly softer than the hypothesized spectrum of astrophysical neutrinos. 
The cosmic ray showering process gives these neutrinos a characteristic distribution in direction, peaked near the observer's horizon, because of the different profiles of atmospheric density the air showers encounter. 

The prompt atmospheric neutrinos are less well understood, as they have not yet been observed experimentally, and the theoretical predictions depend on understanding heavy quark production in cosmic ray-air collisions at high energies. 
Multiple calculations exist \cite{ref:TIG_prompt,ref:MRS_prompt,ref:ERS}, and here we choose the phenomenological ERS estimate of the flux \cite{ref:ERS}, again applying corrections for the input cosmic ray spectrum. 
This model has a normalization uncertainty of about a factor of two, and other calculations predict substantially larger or smaller fluxes. 
Like the conventional atmospheric neutrinos, the energy spectrum of the prompt component arises from the spectrum of the cosmic rays.
However, since the intermediate mesons involved decay so rapidly (with a mean lifetime of $1.04 \times 10^{-12}$ s for the $\mathrm{D}^{\pm}$ at rest, as opposed to $2.60 \times 10^{-8}$ s for the $\mathrm{\pi}^{\pm}$ or $1.24 \times 10^{-8}$ s for the $\mathrm{K}^{\pm}$), losses via interactions are suppressed and the spectrum remains similar to $E^{-2.7}$, and likewise remains essentially isotropic. 

%Overview of fitting procedure (systematic effects as nuisance parameters)
To fit the observed data, we implement the binned Poisson profile likelihood construction described in \cite{ref:IC40}. 
Here, the expected event rates for each flux component are computed by weighting a generalized simulation of neutrinos traversing the Earth and interacting at IceCube according to the model's input neutrino flux.
Comparisons are made in each bin to the observed data. 
For this study, the data are binned in both the reconstructed zenith angle and the energy proxy. 
The main parameter of interest for this fit is the normalization assigned to the astrophysical flux component, while the normalizations of the background components are treated as nuisance parameters. 
Additional nuisance parameters include the difference between the true slope of the cosmic ray spectrum and the assumed model, the efficiency with which the IceCube hardware detects photons emitted in the ice, and the relative contributions to the conventional atmospheric neutrino flux from kaon decays rather than pion decays. 
The nuisance parameters can be constrained using prior information from external sources, and the priors used in this analysis are listed in the fourth column of Table \ref{tab:fit_results}.

\begin{figure}
	\includegraphics[scale=0.7]{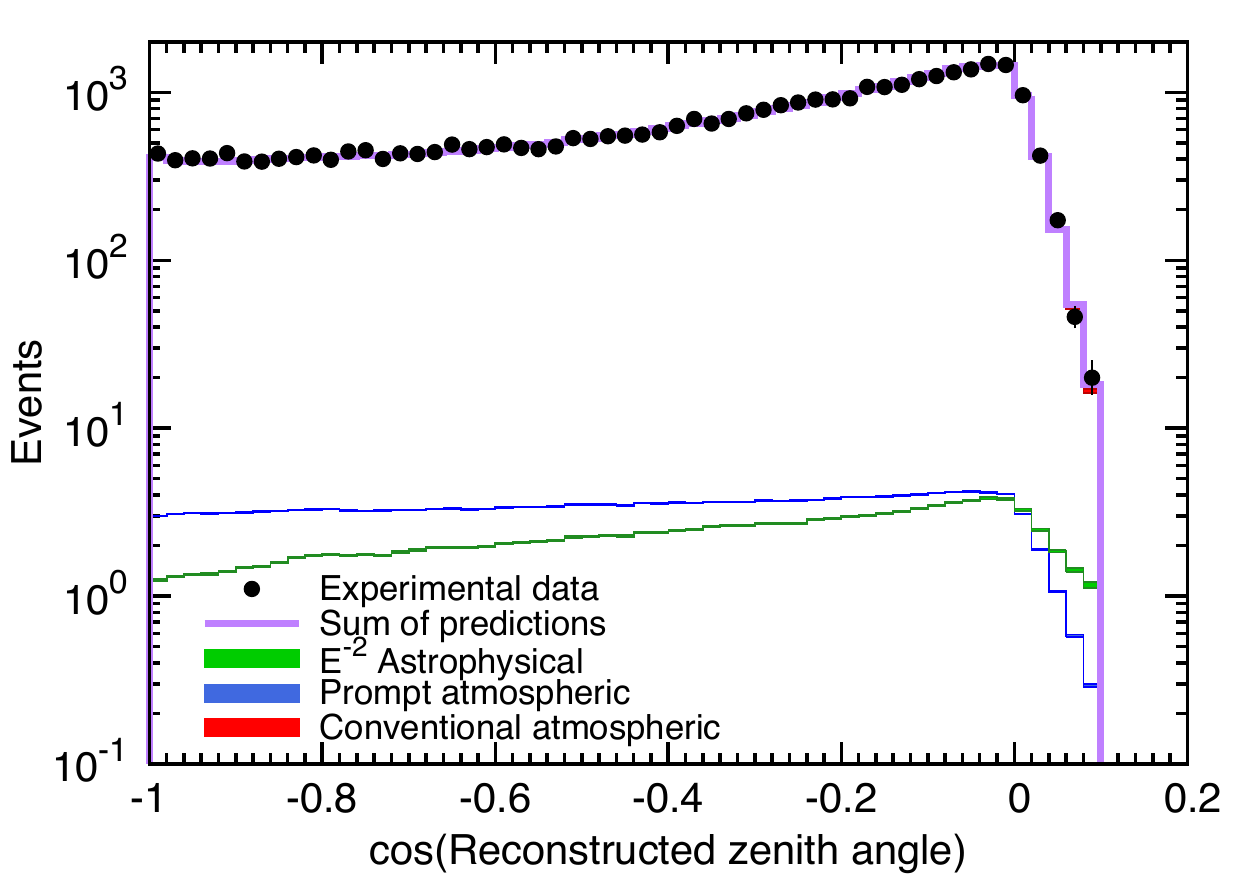}
	\caption{
		The distribution of reconstructed zenith angles of events in the final sample, compared to the expected distributions for the fit of an $E^{-2}$ astrophysical neutrino spectrum. Only statistical errors are shown, though in almost all bins they are small enough to be hidden by the data markers. 
	}
	\label{fig:zenith}
\end{figure}

\begin{figure}
	\includegraphics[scale=0.7]{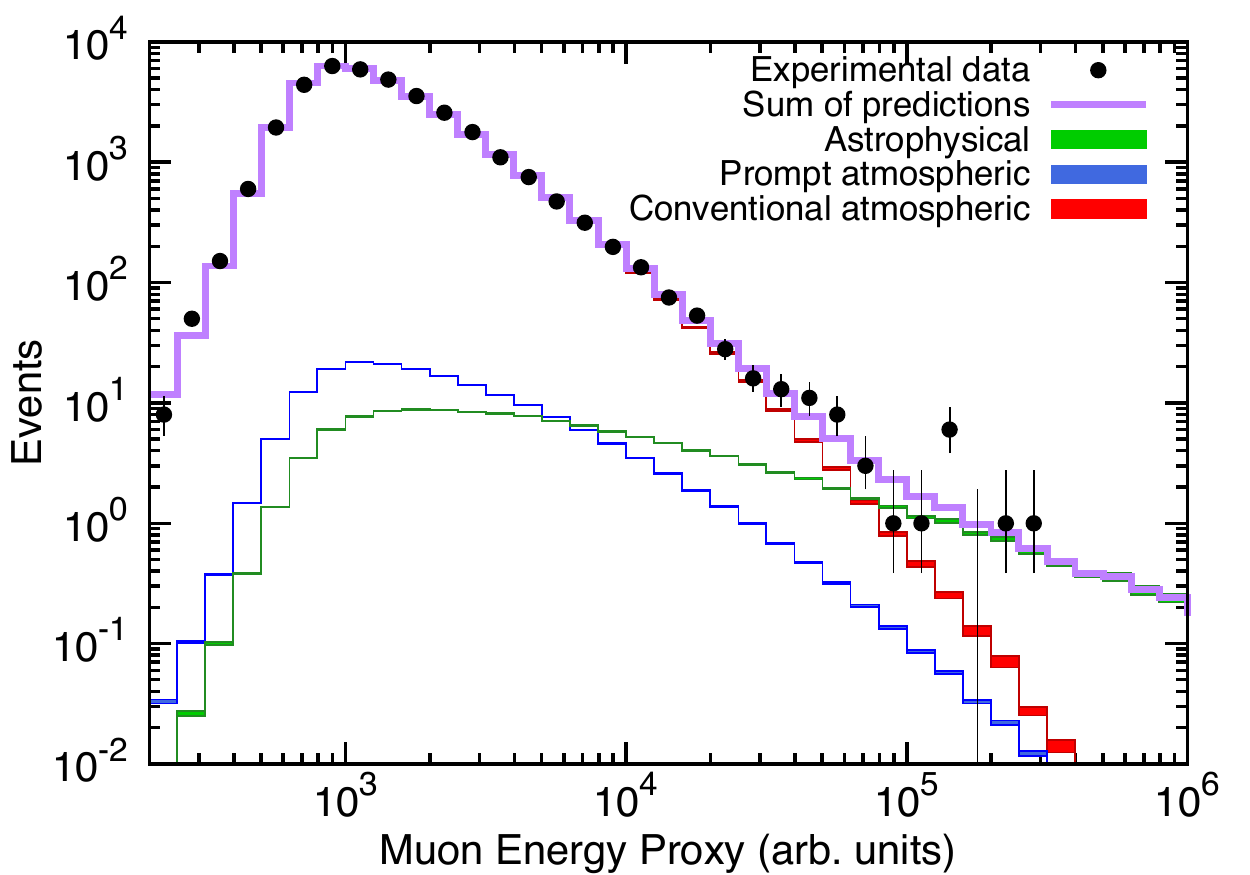}
	\caption{
		The distribution of reconstructed muon energy proxy for events in the final sample, compared to the expected distributions for the fit of an $E^{-2}$ astrophysical neutrino spectrum. Only statistical errors are shown. The energy proxy does not have a linear relationship to actual muon energy, but values $\sim 3 \times 10^3$ are roughly equivalent to the same quantity in GeV. Larger proxy values increasingly tend to underestimate muon energies, while smaller values tend to overestimate. 
	}
	\label{fig:energy}
\end{figure}

\begin{table*}
	\begin{tabular}{lllc}
		\hline \\[-2ex] 
		Parameter & $E^{-2}$ Fit & Best Fit & Prior \\
		\hline \\[-1.5ex] 
		Astrophysical flux normalization per flavor & $9.9^{+3.9}_{-3.4} \times 10^{-19}$ & $1.7^{+0.6}_{-0.8} \times 10^{-18}$ & $\geq0$ \\
		Astrophysical flux index & fixed to 2 & $2.2^{+0.2}_{-0.2}$ & none \\
		[1ex] \hline \\[-1.5ex] 
		HKKMS07 normalization & $0.93^{+0.05}_{-0.04}$ & $0.93^{+0.04}_{-0.04}$ & $\geq0$ \\
		ERS normalization & $0.94^{+1.50}_{-0.94}$ & $0^{+1.05}$ & $\geq0$ \\
		Cosmic ray spectral index change & $-0.024^{+0.011}_{-0.011}$ & $-0.023^{+.001}_{-.0008}$ & $0\pm0.05$ \\ % H3a pop 1: -2.66, with this change -2.632
		Detector optical efficiency & $+9.1^{+0.5}_{-0.5}\%$ & $+9.1^{+0.5}_{-0.5}\%$ & none \\ % Unshadowed fraction = 1.0656
		Kaon production normalization & $1.15^{+0.08}_{-0.07}$ & $1.15^{+0.08}_{-0.07}$ & $1\pm0.1$ \\
	\end{tabular}
	\caption{
		Fit parameters are shown for two case: when an $E^{-2}$ astrophysical flux  with equal flavor composition and equal neutrino and antineutrino components is assumed ($E^{-2}$ Fit), and when the index of the astrophysical flux is allowed to vary (Best Fit). The listed error ranges are 68\% confidence intervals. The gaussian priors are shown as the mean value $\pm$ the standard deviation, but the fit results do not depend substantially on the priors. Units for the astrophysical flux normalization are ${\rm {GeV^{-1} \, cm^{-2} \, sr^{-1} \, s^{-1}} }$, and HKKMS07 \cite{ref:Honda} and ERS \cite{ref:ERS} are the reference conventional and prompt atmospheric fluxes, respectively. 
	}
	\label{tab:fit_results}
\end{table*}

%\textit{Results} ---
%Best fit to data with $E^-2$ and $E^{-\gamma}$, state significance for E-2
The parameter values from fitting 659.5 days of detector livetime using the benchmark set of fluxes are summarized in Tab. \ref{tab:fit_results}, and the projections of the observed and fitted spectra into the reconstructed zenith angle and muon energy proxy are shown in Fig.~\ref{fig:zenith}, and Fig. \ref{fig:energy}, respectively. 
The uncertainties shown for the fit parameters include both statistical and systematic contributions (at the 68\% confidence level), via the profile likelihood, using the ${\chi}^2$ approximation \cite{ref:Wilks}. 
Note that the data point in Fig.~\ref{fig:energy} at muon energy proxy values of around $1.4\times10^5$ should not be taken as an indication of a spectral feature: A fluctuation of this size is expected to occur in approximately 9\% of experiments due to statistical fluctuations, and even a delta function component in the true neutrino spectrum would be broadened into a far wider peak in the muon energy proxy \cite{Note1}. 

The best fit for the astrophysical component is a flux $\Phi(E_{\nu}) = 9.9^{+3.9}_{-3.4} \times 10^{-19}\, \mathrm{GeV}^{-1}\, \mathrm{cm}^{-2}\, \mathrm{sr}^{-1}\, \mathrm{s}^{-1} \left({E_{\nu} \over 100 \mathrm{TeV}}\right)^{-2}$ per flavor. 
The best fit prompt component is  0.94 times the benchmark flux, but is consistent with zero. 
The significance of the non-zero astrophysical flux is evaluated by a likelihood ratio test to the null hypothesis that only atmospheric neutrino fluxes are present, in which case the fitted prompt atmospheric normalization rises to 4.0 times the ERS model. 
An ensemble of trials is used to establish the distribution of the likelihood ratio test statistic, yielding a p-value of $1.1 \times 10^{-4}$ or a single-sided significance of $3.7 \sigma$. 

The range of neutrino energies in which this astrophysical flux is constrained by the data is calculated to be 330 TeV-1.4 PeV. 
The endpoints of this range are found by applying a hard cutoff to one end of the astrophysical flux template, refitting the data with the other astrophysical flux parameters held constant, moving the cutoff inward until the resulting fit likelihood is $0.5 \sigma$ worse than the best fit. 
This gives a conservative estimate of the energy range in which the astrophysical flux is necessary to explain the observed data, although the flux may actually have a greater extent
\footnote{
It should be noted that the method of this calculation differs from that used in \cite{ref:IC59}.
There, the calculation was of the energy range in which the analysis would theoretically be sensitive to a flux, which is independent of the observed data. 
The equivalent sensitive range of this analysis using the technique in~\cite{ref:IC59} is found to be 212 TeV-64.5 PeV. 
}. 
The flux should not be interpreted as existing strictly within this energy range; were this the case simulation trials suggest that this analysis would measure a flux normalization only 5-20\% of the result shown in Table~\ref{tab:fit_results}. 

\begin{figure}
	\includegraphics[scale=.7]{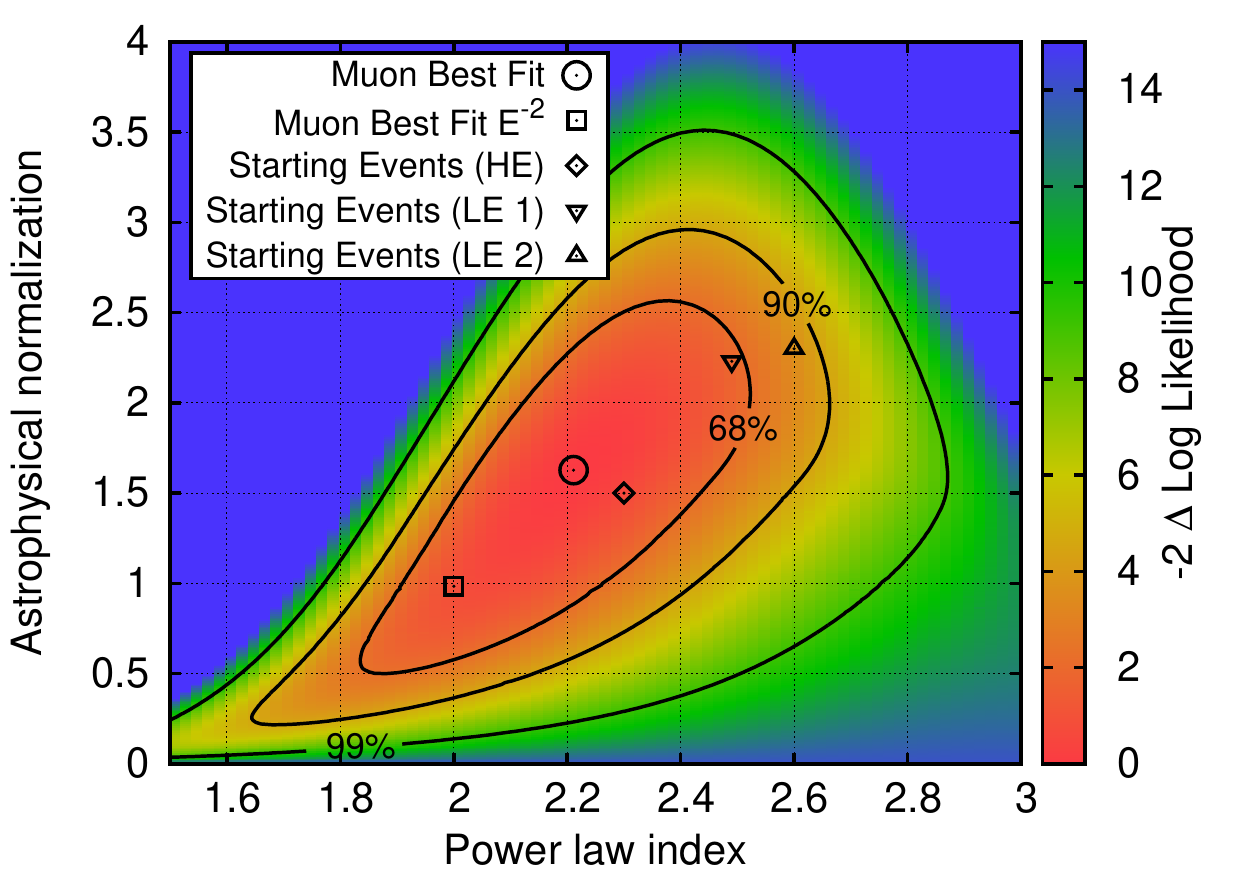}
	\caption{
		Likelihood profile of the astrophysical flux power-law index and the flux normalization at $100\, \mathrm{TeV}$ in units of $10^{-18}\, \mathrm{GeV}^{-1}\, \mathrm{cm}^{-2}\, \mathrm{sr}^{-1}\, \mathrm{s}^{-1}$. While the $E^{-2}$ result is well within the 68\% contour, it is not the overall best fit. 
		Also shown are the best fits from various IceCube analyses of starting events, which generally have good agreement: Starting Events (HE) \protect{\cite{ref:HESE2}}, Starting Events (LE 1) \protect{\cite{ref:JvS}}, Starting Events (LE 2) \protect{\cite{ref:GB}}. 
	}
	\label{fig:freeIndex}
\end{figure}

Since the true flux need not have a spectral index of exactly 2, the fit was repeated allowing the index to vary, leading to a result of $\Phi(E_{\nu}) = 1.7^{+0.6}_{-0.8} \times 10^{-18}\, \mathrm{GeV}^{-1}\, \mathrm{cm}^{-2}\, \mathrm{sr}^{-1}\, \mathrm{s}^{-1} \left({E_{\nu} \over 100 \mathrm{TeV}}\right)^{-2.2 \pm 0.2}$. 
The nuisance parameters do not change significantly except the prompt atmospheric normalization, which falls to zero, as shown in Tab. \ref{tab:fit_results}. 
Figure \ref{fig:freeIndex} shows the confidence regions for the astrophysical flux normalization and spectral index, and compares this result to three other IceCube analyses using starting events \cite{ref:HESE2, ref:JvS, ref:GB}. 
The compatibility of these results is noteworthy because this work uses an independent set of data from the others (a single, near-horizontal, high energy track event is shared with the other samples), while the starting event results are highly correlated with each other. 
The spectral indices found by this work and by the starting event analyses are consistent within their respective uncertainties, but the best fit spectrum for this data set is slightly harder than those for the starting event analyses, particularly those extending to lower energies, which are uniquely able to probe non-atmospheric contributions to the neutrino flux.
A single power law provides an acceptable fit to all data, however, the present data cannot yet rule out the possibility that the astrophysical neutrino flux is not isotropic or that the spectrum is not a pure power law. 

\begin{figure}
	\includegraphics[scale=0.48]{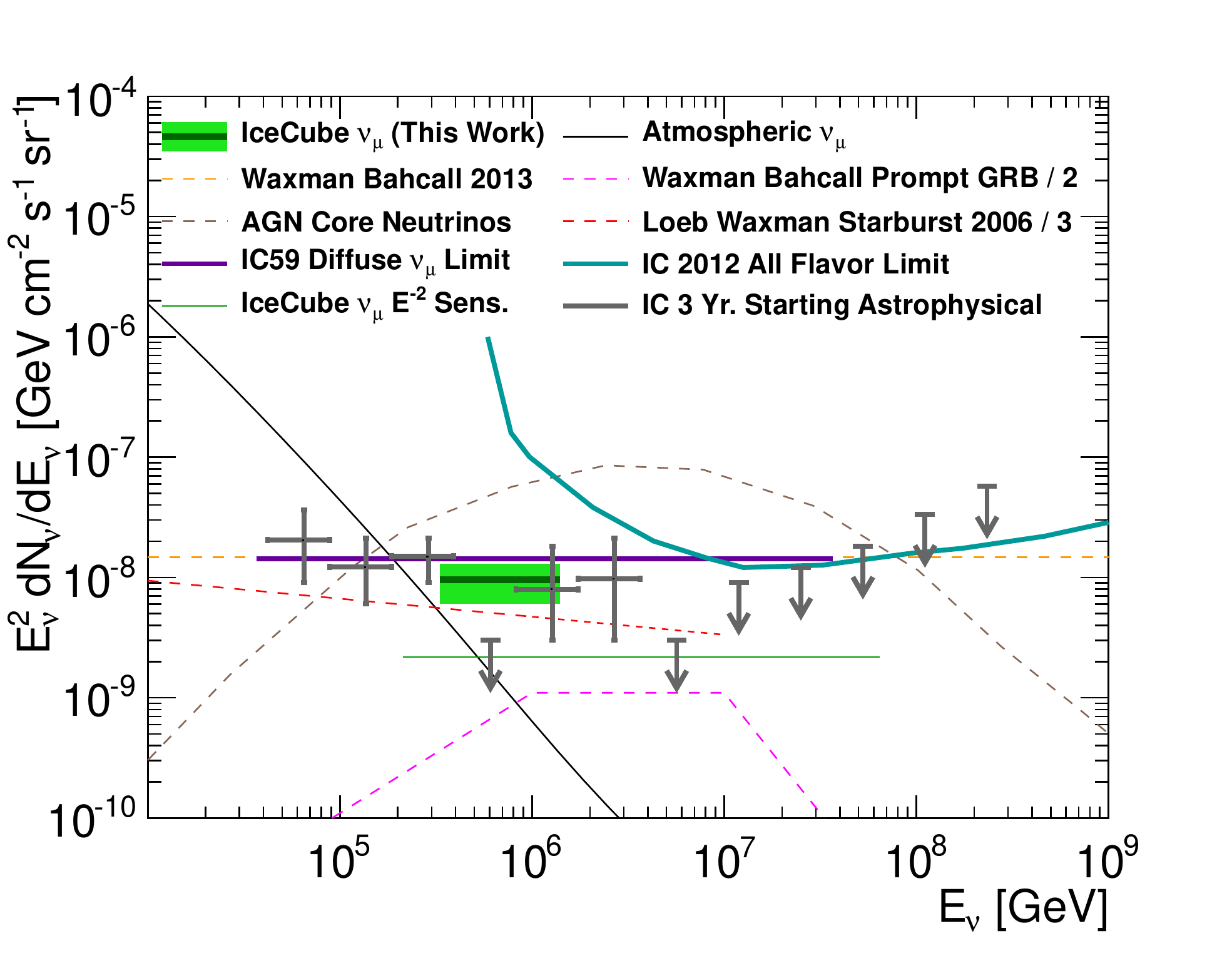}
	\caption{
		Comparison of the best fit per-flavor astrophysical flux spectrum of $E^{-2}$ from this work, assuming a flavor ratio of 1:1:1, (shown in dark green with the 68\% error range in lighter green) to other selected IceCube measurements (heavy lines) \protect{\cite{ref:IC59,ref:HESE2}} and theoretical model predictions (thin, dashed lines) \protect{\cite{ref:Honda,ref:ERS,ref:WB_GRB,ref:WB2013,ref:Stecker2005,ref:LoebWaxman}}. The sensitivity of this analysis is also shown as the thin, green line. 
	}
	\label{fig:fluxes}
\end{figure}

%\textit{Conclusions} ---
%Complementarity to HESE search, similarity of results
In this study we see a clear excess of data above the expected atmospheric neutrino backgrounds at high energies, similar to the result of \cite{ref:HESE2}. 
In particular, despite the fact that these are almost entirely disjoint datasets (a single, near-horizontal track event, event 5 from \cite{ref:HESE2}, appears in both samples), both excesses are consistent in normalization within uncertainties, assuming an $E^{-2}$ spectrum: $9.5\pm3\times10^{-19} {\rm {GeV^{-1} \, cm^{-2} \, sr^{-1} \, s^{-1}} }$ from the starting event study and $9.9^{+3.9}_{-3.4}\times10^{-19} {\rm {GeV^{-1} \, cm^{-2} \, sr^{-1} \, s^{-1}} }$ from this work.
These measurements do use different calculations of the neutrino-nucleon cross-sections, which influence the conversion of the flux into a rate of observed events: 
The starting event study used the calculation of \cite{ref:CSS}, while this study uses the updated calculation from \cite{ref:CSMS}, which differs by 5-10\% at the energies relevant to these analyses, but this is a relatively small effect compared to the uncertainties of these results. 
Thus, the observed data are found to be consistent with a flux consisting of equal parts of all neutrino flavors. 
Similar consistency is seen in a recent analysis of starting events \cite{ref:GB}. 
As shown in Fig.~\ref{fig:freeIndex}, the results for arbitrary power laws are also in good agreement. 
These two measurements are compared in Fig.~\ref{fig:fluxes}, along with other recent measurements and theoretical models. 
The result of this study also suggests that astrophysical neutrinos are present at the several hundred TeV energies where observations were lacking in the dataset of \cite{ref:HESE2}, suggesting that this was merely a statistical fluctuation. 

%Discussion of models
Models of the astrophysical neutrino flux besides unbroken power laws can also be considered. 
Here we examine a small number of representative models. 
One candidate source type is the cores of active galactic nuclei (AGN) \cite{ref:Stecker2005, ref:NMB_AGN, ref:AMM_AGN, ref:MSB_AGN, ref:RM_AGN}. 
A fit of the AGN flux model~\cite{ref:Stecker2005} to the data in this analysis demonstrates in an incompatibility in the normalization, with the predicted flux being too large by a factor of 6.
Another possible source class are regions with high star formation including Starburst galaxies \cite{ref:LoebWaxman, PhysRevD.87.063011, PhysRevD.89.127304, 0004-637X-806-1-24, PhysRevD.89.083004, Chakraborty201535}. 
Comparing the $E^{-2.15}$ spectrum proposed by \cite{ref:LoebWaxman} to the data reported here, we find that it is compatible after its normalization is multiplied by a factor of 2.5. 
Finally gamma ray bursts (GRBs) have been long considered candidates for neutrino production \cite{ref:WB_GRB, PhysRevLett.111.121102, PhysRevLett.95.061103, PhysRevD.73.063002, Baerwald2012508}, but recent dedicated searches by IceCube for neutrinos correlated with GRBs have placed strong limits disfavoring this hypothesis \cite{ref:IC_GRB}. 

%Inference on number of sources based on PS limits
While this work represents the first strong evidence for an astrophysical $\nu_\mu$ flux in the Northern Hemisphere, the sources producing these neutrinos remain unknown. 
Although muon events in IceCube have sub-degree angular resolution, recent IceCube searches for point-like and extended sources of muon neutrinos found no statistically-significant evidence for event clustering, nor correlation of neutrinos with known astrophysical objects~\cite{ref:IC86PSPaper}. 
In the Northern Hemisphere, the point source flux upper limits are $10-100$ times lower than the total diffuse flux level observed here, so the flux cannot originate from a small number of sources without violating those limits.  
The constraint on the number of sources was explored with a simple simulation where sources were injected uniformly over the Northern sky, with fluxes at the maximum levels allowed by the point source upper limit at each selected point, until the total flux reached the measured diffuse flux. 
On average, at least $70$ sources are required to maintain consistency with the point source upper limits. 
This assumes each source is a true point source and emits an unbroken $E^{-2.2}$ power-law flux.
If the sources instead follow harder $E^{-2}$ power law spectra, the diffuse flux could be split across an average of $\sim40$ sources while remaining consistent with the point source analysis. 
Given that the diffuse flux in the Southern Hemisphere is observed at a similar flux level, this observation suggests that the flux has a large isotropic component dominated by a large population of extragalactic sources, although local sources can still have significant contributions. 

\begin{acknowledgments}

We acknowledge the support from the following agencies:
U.S. National Science Foundation-Office of Polar Programs,
U.S. National Science Foundation-Physics Division,
University of Wisconsin Alumni Research Foundation,
the Grid Laboratory Of Wisconsin (GLOW) grid infrastructure at the University of Wisconsin - Madison, the Open Science Grid (OSG) grid infrastructure;
U.S. Department of Energy, and National Energy Research Scientific Computing Center,
the Louisiana Optical Network Initiative (LONI) grid computing resources;
Natural Sciences and Engineering Research Council of Canada,
WestGrid and Compute/Calcul Canada;
Swedish Research Council,
Swedish Polar Research Secretariat,
Swedish National Infrastructure for Computing (SNIC),
and Knut and Alice Wallenberg Foundation, Sweden;
German Ministry for Education and Research (BMBF),
Deutsche Forschungsgemeinschaft (DFG),
Helmholtz Alliance for Astroparticle Physics (HAP),
Research Department of Plasmas with Complex Interactions (Bochum), Germany;
Fund for Scientific Research (FNRS-FWO),
FWO Odysseus programme,
Flanders Institute to encourage scientific and technological research in industry (IWT),
Belgian Federal Science Policy Office (Belspo);
University of Oxford, United Kingdom;
Marsden Fund, New Zealand;
Australian Research Council;
Japan Society for Promotion of Science (JSPS);
the Swiss National Science Foundation (SNSF), Switzerland;
National Research Foundation of Korea (NRF);
Danish National Research Foundation, Denmark (DNRF)

\end{acknowledgments}

\bibliography{paper}

\newpage

\ifx \standalonesupplemental\undefined
\setcounter{page}{1}
\setcounter{figure}{0}
\setcounter{table}{0}
\fi

\newcolumntype{L}[1]{>{\arraybackslash}p{#1}}
\newcolumntype{C}[1]{>{\centering\arraybackslash}p{#1}}
\newcolumntype{R}[1]{>{\hfill\arraybackslash}p{#1}}

\renewcommand{\thepage}{Supplementary Methods and Tables -- S\arabic{page}}
\renewcommand{\figurename}{SUPPL. FIG.}
\renewcommand{\tablename}{SUPPL. TABLE}

%\title{Supplemental Material}

\section*{Data Selection}

This section describes in greater detail the quality criteria applied to select the events used in this study. 
These criteria were designed by comparing 805.5 hours of experimental data from the 2010 data-taking period (representing 10\% of the data from that period and 5\% of the total data used in the full analysis) to simulated data. 
The simulation included a cosmic-ray dataset equivalent to approximately 264 hours of air shower background
%with a spectrum as given by \cite{ref:polygonato}
and a neutrino dataset weighted to both a conventional atmospheric neutrino spectrum and a hypothetical $E^{-2}$ powerlaw spectrum (representing a possible astrophysical flux). 

\begin{enumerate}
	\item Muon Filter: The data available as input for this analysis are those selected by the IceCube detector's `online' filters, which determine the subset of the data to be transmitted to the Northern Hemisphere. 
	The filters operate on the output of the detector data acquisition system, which generates event readouts whenever a trigger condition is met. 
	Several triggers are implemented, but a representative example is the simple multiplicity trigger, whose condition is that some minimum number of detector modules (typically 8) observe light within a short time window (typically 5 microseconds), where each of the contributing modules must have made its observation within one microsecond of one of its four vertically neighboring modules. 
	The triggers are designed to read out the detector when useful physics signals are likely to be present, which the filters then attempt to reconstruct and classify. 
	The `Muon Filter' is designed to select events compatible with neutrino-induced muons passing through the detector. This includes upwards-going events as well as high-energy downwards-going events. Low-energy downwards-going events are mostly air shower muons which are uninteresting for this analysis. 
	Event energies are estimated very simplistically at this level, using the total amount of observed light as a proxy. 
	
	\item Data Reduction: Detailed reconstructions are performed on the data selected by the Muon Filter, which can reveal that the event should not have originally passed the filter criteria, so reapplication of the filter criteria can remove many of the air shower muons which are uninteresting to this analysis. 
	Additionally, a correct track reconstruction should generally pass closer to the points where more photons were detected, so the average distance from the reconstructed track to the detection points (the detector modules which observed light), weighted by number of photons detected, should be small. 
	Events with up-going reconstructions whose weighted distance was greater than 200 meters were therefore rejected, unless they were reasonably bright (more than 100 photoelectrons observed). 
	Some of the air shower events are actually coincident, with bundles of muons from unrelated showers passing through the detector during the same readout window. 
	A separation algorithm is also applied to attempt to separate these into distinct sub-events, which are then reconstructed normally. 
	
	\item Zenith Angle: The main power of this selection to reject air shower muon bundles derives from selecting events which pass through considerable overburden to reach the detector. At some zenith angle, $\theta$, the overburden is no longer sufficient and air shower events will reach the detector at a non-negligible rate. 
	Events with $\cos(\theta) < 0.1$ ($\theta \approx 85^\circ$) must pass through at least 12 km of water equivalent to reach the top of the IceCube detector, so only these events are selected for use in this analysis. 
	
	\item Track Quality: In most of the zenith angle range used by this analysis, the Earth blocks cosmic ray muons entirely from reaching the detector. 
	Some such events are present in this part of the observable space in the reconstructed data, but only because their reconstructions are inaccurate. 
	Likewise, neutrino-induced events whose directions are severely misreconstructed need to be rejected. 
	Two methods are used here, both of which derive from the fact that the fit used to reconstruct these data is a maximum likelihood calculation, and additional information can be extracted from the likelihood description besides the parameter values of the best fit. 
	The first variable used, known as the `paraboloid sigma' uses the shape of the profile likelihood around the best-fit point to estimate the statistical uncertainty on the location of that point; the second, the `reduced log likelihood' (RLogL) attempts to use the best obtained likelihood value as a global measure of the success of the fit. 
	The combined cut on these variables is shown in Suppl. Fig.~\ref{fig:track_quality_cut}. 
	\label{itm:track_quality_cut}
	
\begin{figure}
	\centering
	\subfloat[0.48\textwidth][Relative rate of simulated air shower events (arb. units)]{
		\includegraphics[width=7.5cm]{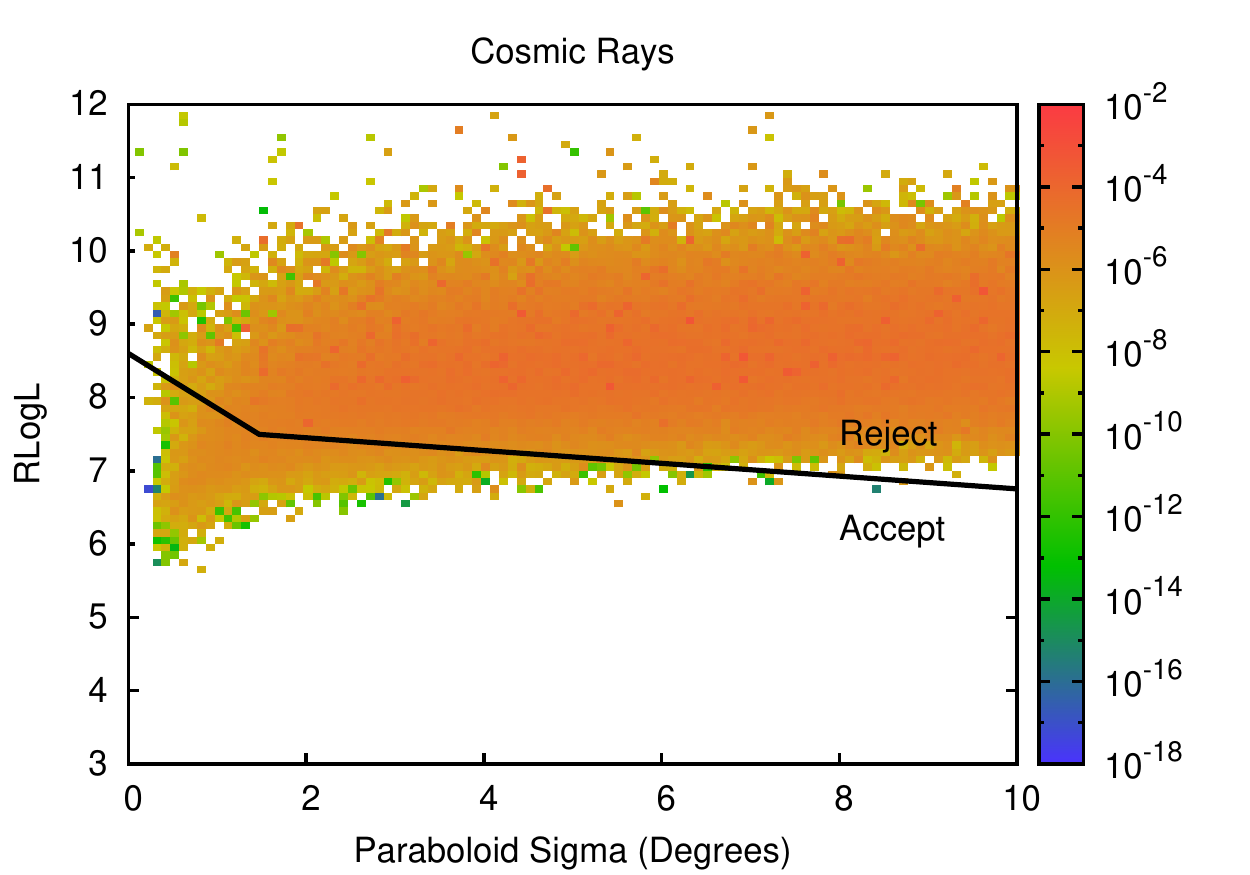}
	}
	
	\subfloat[0.48\textwidth][Relative rate of simulated $E^{-2}$ neutrino events (arb. units)]{
		\includegraphics[width=7.5cm]{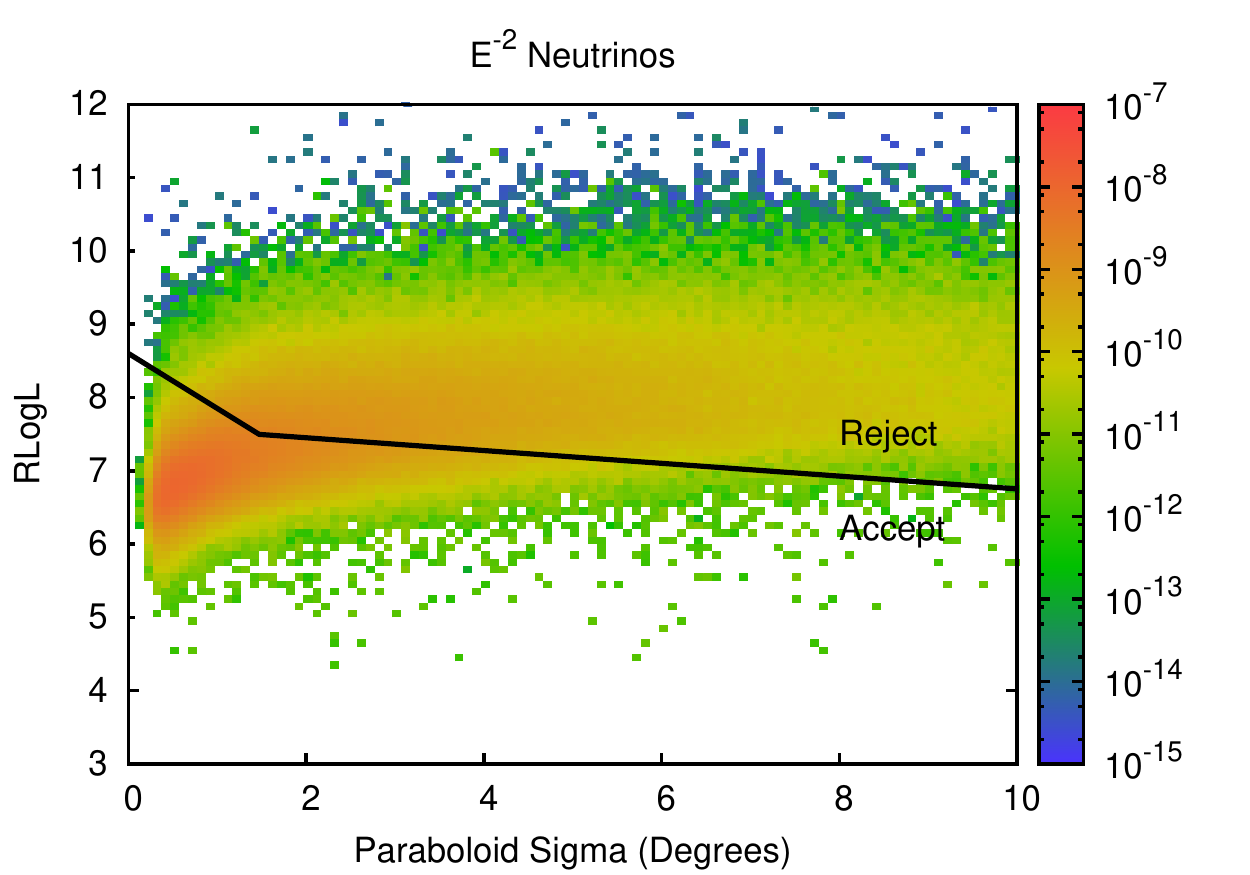}
	}
	\caption{The distributions of simulated air shower background events and $E^{-2}$ signal neutrino events in two measures of directional reconstruction quality, whose meanings are described in the text (Data Selection \autoref{itm:track_quality_cut}). Many air shower events remain in the sample at this stage only because their reconstructed directions are incorrect, so these variables allow separating these types of data. }
	\label{fig:track_quality_cut}
\end{figure}

	\item Bayesian Reconstruction: Another means of eliminating misreconstructed events is to compare the unconstrained track reconstruction with one which has incorporated a bayesian prior that the majority of observed events are truly down-going, and should be reconstructed as such. 
	In the case that an event is correctly reconstructed, the evidence due to the data should overwhelm this prior, and both reconstructions will have similar results. 
	In the up-going region, one can make the simplifying assumption that if the unconstrained reconstruction performed better by some constant factor (corresponding to the floor in the prior) than the constrained reconstruction, it was probably trustworthy. 
	For events in the down-going region which are correctly reconstructed, the reconstructions with and without the prior should be similar, and thus the difference in their log likelihoods will be related by the prior, so the distribution of well-reconstructed events should approximately trace out the prior as a function of zenith angle. 
	The cut which was used on this variable is shown in Suppl. Fig.~\ref{fig:bayes_cut}. 
	
\begin{figure}
	\centering
	\subfloat[0.48\textwidth][Relative rate of simulated air shower events (arb. units)]{
		\includegraphics[width=7.5cm]{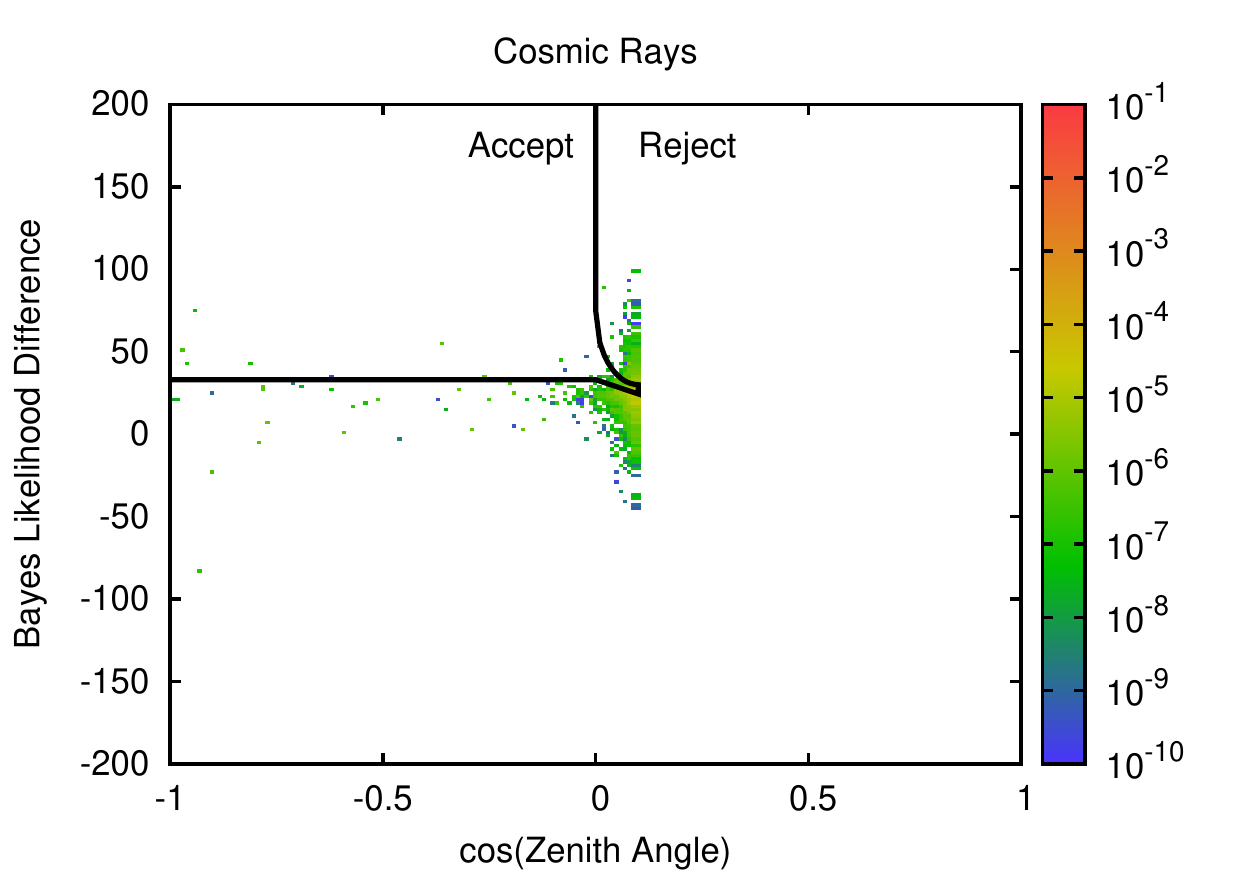}
	}
	
	\subfloat[0.48\textwidth][Relative rate of simulated $E^{-2}$ neutrino events (arb. units)]{
		\includegraphics[width=7.5cm]{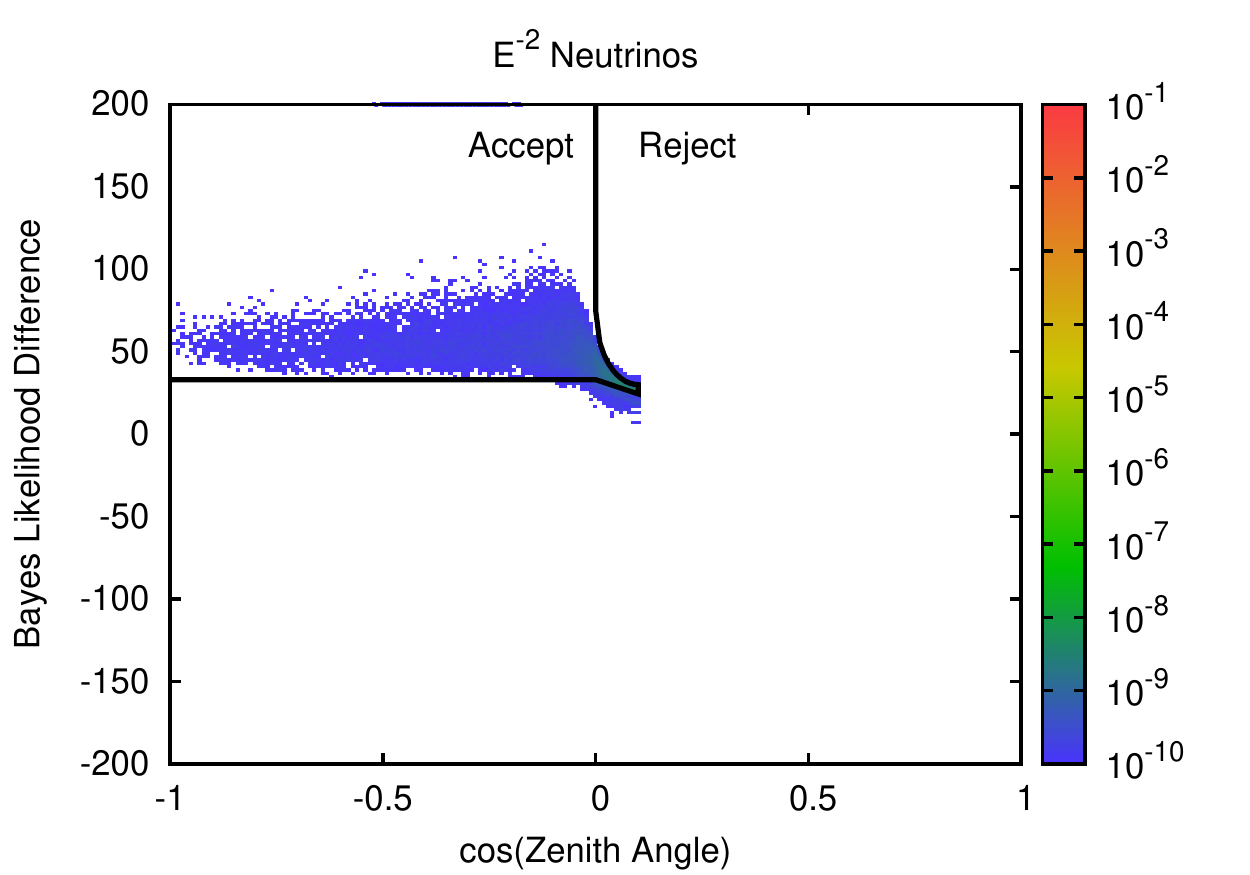}
	}
	\caption{The distribution of simulated air shower background events and $E^{-2}$ signal neutrino events as a function of reconstructed zenith angle and difference of log likelihoods between reconstructions which do and don't contain a bayesian prior that the events should be reconstructed down-going according to an approximate, known distribution of background event angles. }
	\label{fig:bayes_cut}
\end{figure}
	
	\item Zenith/Brightness: After the bayesian reconstruction criterion, the remaining air shower background events should generally be well described by down-going reconstructions, although they are still the dominant component of the data above the horizon. 
	The air showers are typically dim, however, after being attenuated in the overburden, while the muons produced by neutrinos can be produced much closer and so can arrive at the detector with much higher energies. 
	By sacrificing all events with relatively low energies, as simplistically measured by the number of detector modules which receive light, it is possible to eliminate virtually all of the expected air showers at a given zenith angle, while leaving a window for higher energy neutrino events.
	At the same time, very dim events with up-going reconstructions are also eliminated, as this accounts for few neutrinos which are actually valuable to this analysis, but also a few remaining air shower events with poor reconstructions. 
	This cut is shown in Suppl. Fig.~\ref{fig:zenith_nchan_cut}. 
	
\begin{figure}
	\centering
	\subfloat[0.48\textwidth][Relative rate of simulated air shower events (arb. units)]{
		\includegraphics[width=7.5cm]{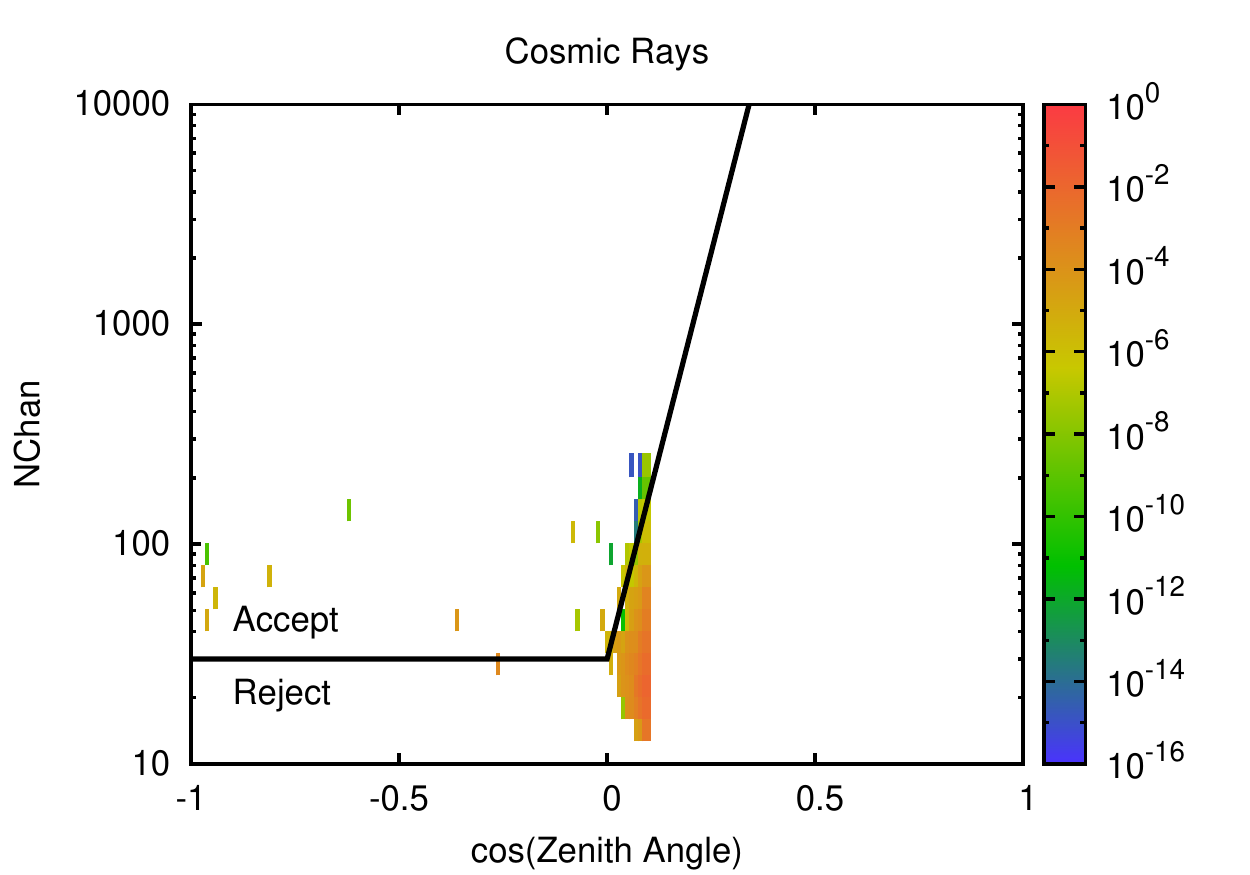}
	}
	
	\subfloat[0.48\textwidth][Relative rate of simulated $E^{-2}$ neutrino events (arb. units)]{
		\includegraphics[width=7.5cm]{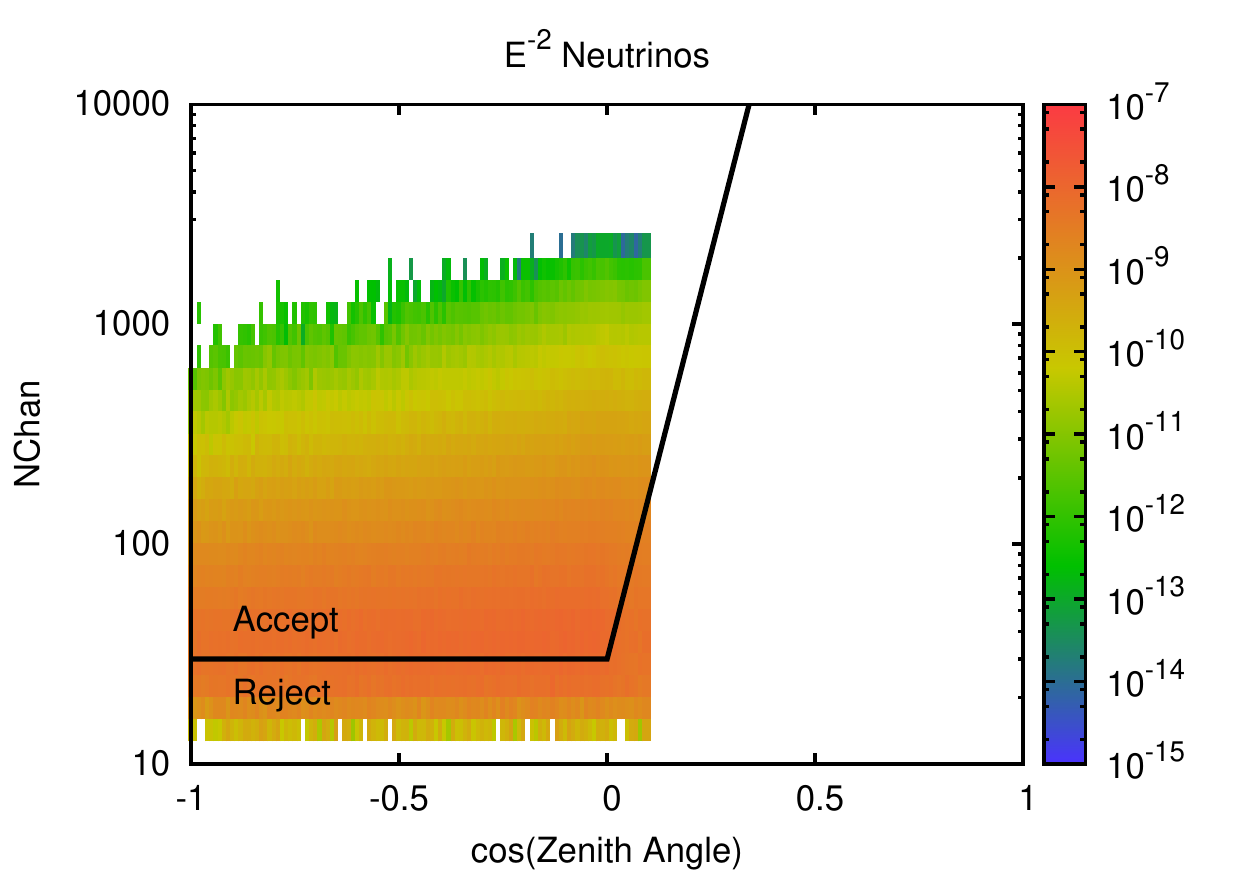}
	}
	\caption{The distribution of simulated air shower background events and $E^{-2}$ signal neutrino events as a function of reconstructed zenith angle and number of detector channels triggered (NChan). The air showers generally have down-going reconstructions and low brightness.}
	\label{fig:zenith_nchan_cut}
\end{figure}
	
	\item Split Event: Despite the separation algorithm applied during the Data Reduction step, some coincident air shower events remain. Therefore, we attempt to remove them by brute force by dividing each event into halves, reconstructing both halves separately, and testing whether the two separate reconstructions give a better fit than a single reconstruction. Specifically, each event is divided in half both at the mean time of light detection, and along a plane perpendicular to the single particle reconstruction and passing through the amplitude-weighted mean position of the light detections. Both division methods are tested, and if either yields a better reconstruction the event is rejected. 
	The fit is here considered better according to the number of photons appearing as unscattered with respect to the track hypotheses, so if the split reconstructions yield more than twice as many unscattered photons as the single reconstruction, we conclude that the event is coincident and remove it. 
	This final cut eliminates a portion of the remaining air shower background events, but is included mostly for safety against contamination unforeseen due to the limited simulation statistics used in developing the selection. 
\end{enumerate}

Suppl. Tab.~\ref{tab:cut_progression} shows the overall results of the data selection process. Background from cosmic ray air showers is reduced by a factor of approximately $5.8\times10^7$ so that it makes up only about 0.1\% of the final data, while 23.8\% of the neutrinos from a hypothetical $E^{-2}$ flux which trigger the detector are expected to be retained. 

\begin{table*}
	\centering
	\begin{tabular}{|p{4cm}|p{2cm}R{1.6cm}|p{2cm}R{1.6cm}|R{3cm}|}
	\hline
	Cut level & \multicolumn{2}{c|}{Air shower background} & \multicolumn{2}{c|}{Conv. atmos. neutrinos} & $\mathrm{E}^{-2}$ signal neutrinos \\
	& Rate (Hz) & Number in full sample & Rate (Hz) & Number in full sample & Fraction surviving \\
	\hline
	Muon Filter
	& 32.6 & $1.86\times10^{9}$ & $1.17\times10^{-2}$ & 667000 & 0.932 \\
	\hline
	Data Reduction
	& 10.6 & $6.04\times10^{8}$ & $9.13\times10^{-3}$ & 520000 & 0.897 \\
	\hline
	Zenith Angle & 3.88 & $2.21\times10^{8}$ & $8.08\times10^{-3}$ & 460000 & 0.688 \\
	\hline
	Track Quality & $4.59\times10^{-4}$ & 26100 & $1.36\times10^{-3}$ & 77200 & 0.313 \\
	\hline
	Bayesian Reconstruction & $1.66\times10^{-4}$ & 9460 & $1.12\times10^{-3}$ & 63800 & 0.286 \\
	\hline
	Zenith/Brightness & $2.01\times10^{-6}$ & 115 & $5.75\times10^{-4}$ & 32800 & 0.238 \\
	\hline
	Split Event  & $5.56\times10^{-7}$ & 31.7 & $5.75\times10^{-4}$ & 32800 & 0.238 \\
	\hline
	\end{tabular}
	\caption{Rates and fractions of simulated data surviving by type as a function of the level of selection applied. Efficiencies are with respect to the detector trigger. Note that the cuts following the `Data Reduction' do not strictly have a fully defined ordering; the order shown here is simply chosen for presentation purposes.}
	\label{tab:cut_progression}
\end{table*}

\section*{Energy Estimation}

The main power in this analysis to distinguish the signature of astrophysical neutrinos from the background of atmospheric neutrinos is in the different energy spectra of these fluxes. 
However, the neutrino energies cannot be measured directly, so in this work the energies of the muons produced by the neutrinos are reconstructed, using the fact that the average energy loss rate of high energy muons is proportional to the muon energy. 
Furthermore, since the muon tracks are not required to begin within the instrumented volume of the detector, the energy of the muon on arrival to the detector may be arbitrarily much smaller than its initial energy. 
These facts considerably limit the amount of information which can be extracted about the energies of the neutrinos themselves. 

In addition, the extracted energy information is also limited by the practical capabilities of the muon energy reconstruction. 
The energy reconstruction is selected so that the computed proxy has the highest possible resolution (among currently available methods), but it is not necessarily unbiased and it does not have a one-to-one relationship to true muon energies. 
While an ideal reconstruction would have a one-to-one mapping to the true physical parameter, fluctuations such as the number and size of stochastic energy losses, and variations of the position of the muon track within the detector make this impossible to realize. 
Bias of the estimator can be avoided, or at least largely removed by calibration, but this serves no purpose in the context of the forward-folding maximum likelihood fit, since as long as the proxy is related on average to the true parameter by a monotonic function, the particular choice of this function simply alters the experimental data distribution and the simulated template distributions in the same way, and so cancels out of the likelihood. 
In addition, it is not possible to carry out a calibration procedure fully correctly a priori, since the relationship between the true parameters and the reconstructed proxy depends on the true neutrino energy spectrum. 

\begin{figure}
	\centering
	\subfloat[0.48\textwidth][Distribution of energy proxy values arising from different true muon energies. ]{
		\includegraphics[width=7.5cm]{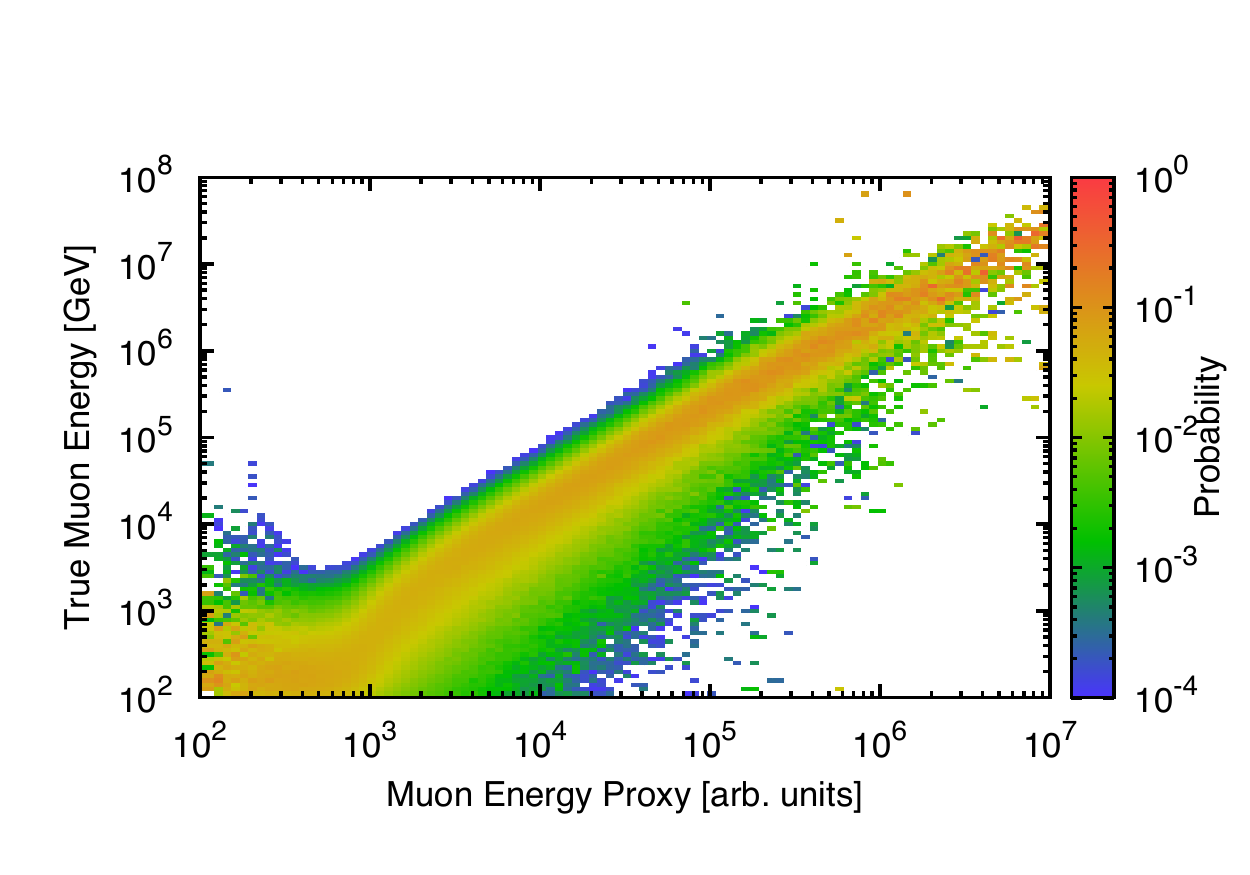}
	}
	
	\subfloat[0.48\textwidth][Distribution of energy proxy values arising from different true neutrino energies.]{
		\includegraphics[width=7.5cm]{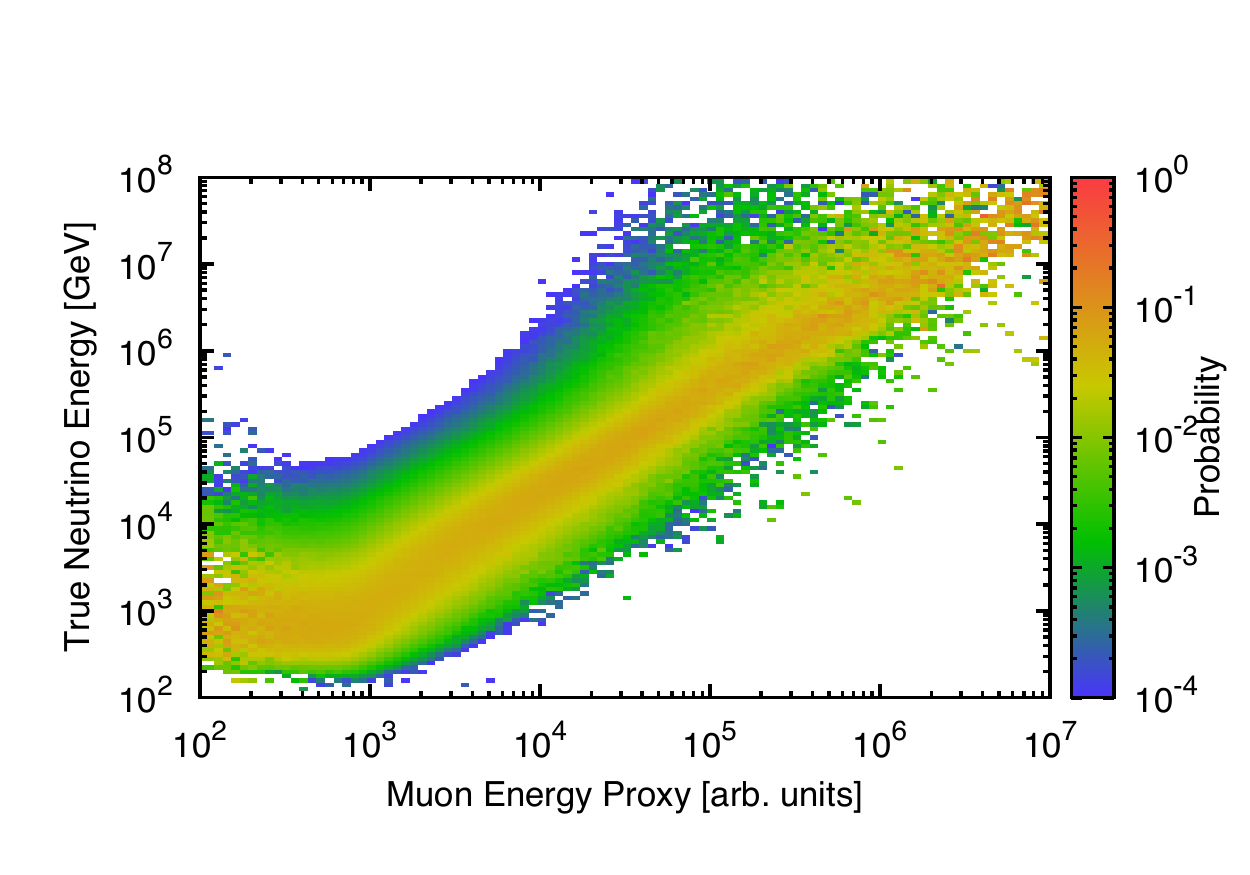}
	}
	\caption{Each column in the figure has been independently normalized to form a PDF of possible true parameter values. The best fit result with an $E^{-2.2}$ astrophysical flux has been assumed. Fluctuations and missing data at the edges of the distributions are due to limited simulation statistics.}
	\label{fig:energy_proxy_from_true}
\end{figure}

While calibration of the energy proxy is not relevant for the maximum likelihood fit, the relationship of the energy proxy to the relevant physical energies is of general interest, and can be explored using the results of the fit. 
Suppl. Fig.~\ref{fig:energy_proxy_from_true} shows the results of weighting a set of simulated neutrino events to the best-fit spectrum produced by the fit and plotting the distributions of the muon energy proxy against the true muon energy at the point of closest approach to the detector center and the true primary neutrino energy. 
Each bin in the energy proxy has been independently normalized, eliminating the influence of the neutrino energy spectrum on the distribution of the proxy, making clear the probability of a proxy value arising from each possible true parameter value. 
The feature which appears at low energy proxy values (and low true particle energy values) is characteristic of the transition to the low muon energy region, where energy loss is dominated by ionization and varies less strongly with energy, from the high muon energy region in which stochastic losses dominate and the energy loss rate varies more rapidly with energy. 
For energy proxy values larger than $\sim10^4$ the most probable true muon energy is somewhat larger (for example, $\sim2.4\times10^5\,\mathrm{GeV}$ for a proxy value of $10^5$). 
The factor relating the energy proxy to the most probable associated neutrino energy is generally larger, for all proxy values, due to the primary neutrino energy being strictly greater---sometimes much so---than the muon energy at the detector (the most probable neutrino energy leading to a proxy value of $10^5$ is $\sim 3.3 \times 10^5\,\mathrm{GeV}$). 

\begin{figure}
	\centering
	\includegraphics[width=7.5cm]{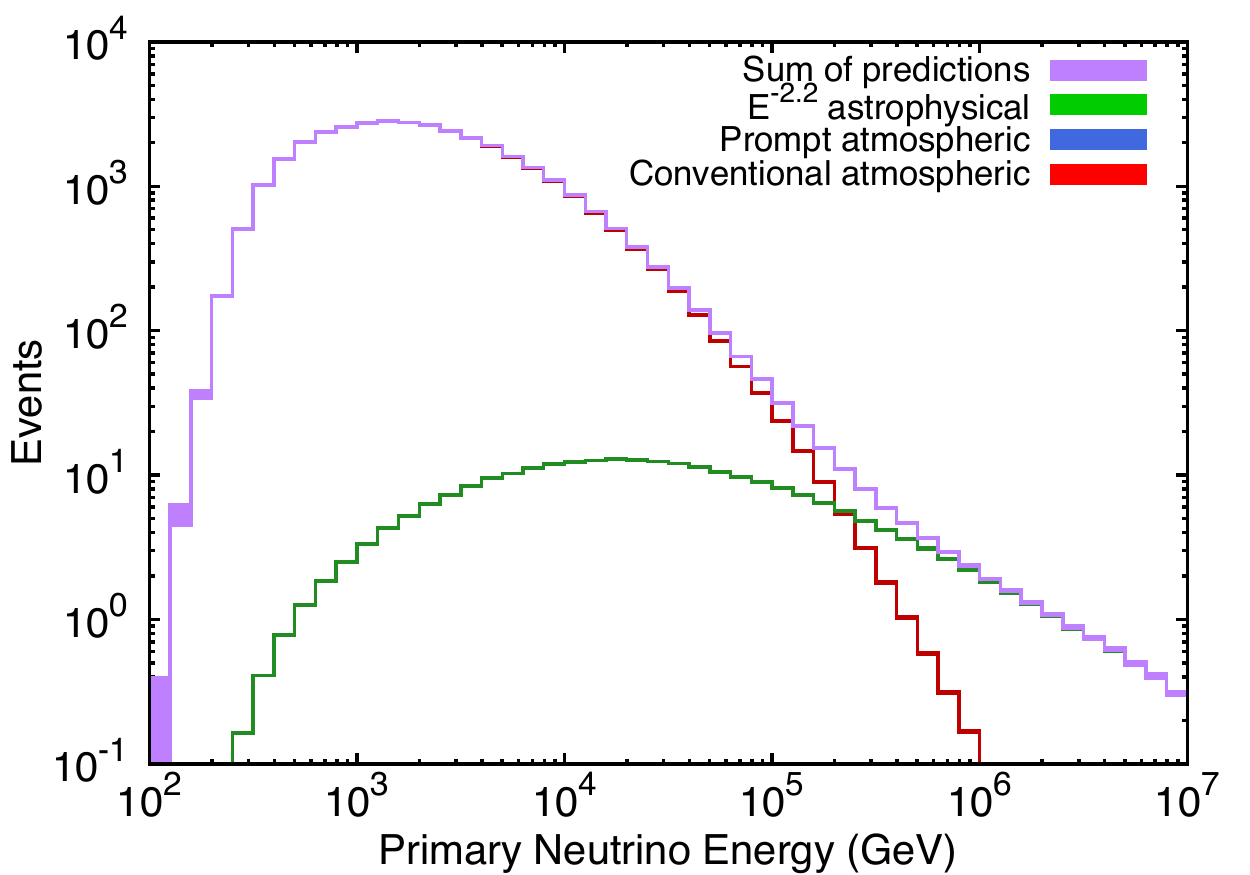}
	\caption{The flux templates obtained for the best fit using a free power law index as functions of primary neutrino energy.}
	\label{fig:nu_en_dist}
\end{figure}

Suppl. Fig.~\ref{fig:nu_en_dist} shows the energy spectra of the fluxes which contribute to the overall best fit. 
Convolving the versions of these distributions and with those of Fig. 4b for an $E^{-2}$ astrophysical neutrino spectrum   yield the curves in Fig. 2 of the main text. 

\clearpage

\end{document}